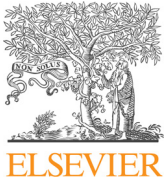
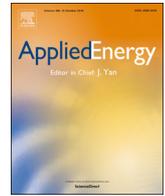

# A data-driven approach for discovering heat load patterns in district heating

Ece Calikus[a],[*], Sławomir Nowaczyk[a], Anita Sant'Anna[a], Henrik Gadd[b,c], Sven Werner[c]

[a] *Center for Applied Intelligent Systems Research, Halmstad University, Sweden*
[b] *Öresundskraft, Helsingborg, Sweden*
[c] *School of Business, Engineering and Science, Halmstad University, Sweden*

## HIGHLIGHTS

- A data-driven approach is proposed to discover heat load patterns in district heating.
- The first large-scale analysis of all the buildings in six different categories is presented.
- We showcase how typical and atypical behaviors look like in the entire network in Sweden.
- The results show that our method has a high potential to be deployed and used in practice.

## ARTICLE INFO

## ABSTRACT

Understanding the heat usage of customers is crucial for effective district heating operations and management. Unfortunately, existing knowledge about customers and their heat load behaviors is quite scarce. Most previous studies are limited to small-scale analyses that are not representative enough to understand the behavior of the overall network. In this work, we propose a data-driven approach that enables large-scale automatic analysis of heat load patterns in district heating networks without requiring prior knowledge. Our method clusters the customer profiles into different groups, extracts their representative patterns, and detects unusual customers whose profiles deviate significantly from the rest of their group. Using our approach, we present the first large-scale, comprehensive analysis of the heat load patterns by conducting a case study on many buildings in six different customer categories connected to two district heating networks in the south of Sweden. The 1222 buildings had a total floor space of 3.4 million square meters and used 1540 TJ heat during 2016. The results show that the proposed method has a high potential to be deployed and used in practice to analyze and understand customers' heat-use habits.

## 1. Introduction

Future energy systems are facing critical challenges such as the steady growth of energy demand, energy resource depletion, and increasing emissions of carbon dioxide ($CO_2$) and other greenhouse gases. District heating (DH) can play a vital role in the implementation of future sustainable energy systems from the renewable [1], European [2], national [3], enhancement [4], and heat recycling [5] perspectives. Implementing these systems will contribute to a decrease in carbon emissions. However, the present generation of district heating technologies must be improved to achieve the target of a 100% renewable energy supply system. The concept of 4th generation district heating [6] discusses how to design efficient and reliable networks and considers environmentally friendly heat production units.

The most important factor in increasing the efficiency in such systems is reducing distribution temperatures so that the quality between the energy supply and demand improves [7]. Achieving low temperatures in the network requires intelligent control systems and elaborated strategies for continuous identification of operation errors causing high return temperatures. To design such strategies, it is crucial to have in-depth knowledge of the customers and a better understanding of their heat use, as even a single substation can have a significant impact on the global efficiency of the system.

Heat load patterns represent the most "typical" behaviors in DH networks and provide information on how different customer groups use heat. Analyzing such patterns is quintessential for effective DH operation and management [8]. This analysis can then be used by DH companies to optimize their operations, to implement new control strategies, and personalize demand management for specific customer groups. Furthermore, this analysis can help decision makers to develop

[*] Corresponding author.
*E-mail address:* ece.calikus@hh.se (E. Calikus).

energy efficiency policies and roadmaps.

Another important aspect is the analysis of DH customers who exhibit abnormal heat use. Even a single problematic customer can influence the overall performance of the network. Traditionally, DH companies have accepted the inefficient heat use and tried to improve operations on the production side. However, this approach is not going to work in 4th generation DH. Up until now, there have been only a few works focusing on the demand side, for example, aiming to reduce peak loads. However, most of them have relied on naive approaches such as using the total head demand or building age as a measure of inefficiency. The biggest challenge is the lack of knowledge about customers and how they use heat since many factors can lead to abnormal heat demand, including poor substation control, unsuitable control strategies, faults, and so forth. Therefore, it is of great interest to identify customers with abnormal heat load profiles or unsuitable control strategies for further investigation.

Discovering typical and abnormal patterns is a complex task, especially for DH systems involving many customers with different characteristics. Heat demand can be affected by several factors [9] such as activities taking place in the buildings, outside temperatures, incident solar radiation, socio-cultural factors, and so on. The knowledge discovered in Sweden may not be directly applicable to, e.g., Italy [8]. Furthermore, inspecting the behaviors of all buildings in the entire network is prohibitively time-consuming. All these reasons make data-driven solutions a necessity for the large-scale analysis of district heating systems. Automating the work allows for more customer personalization, accounts for more factors, and makes it possible to re-do the analysis easily in different parts of the world to discover unique patterns specific to each country or region.

In this work, we present a data-driven approach for automatic heat load pattern discovery and perform the first large-scale analysis of the heat load behaviors in two DH networks. Our contributions are:

1. Heat load pattern discovery: we develop a new method to discover groups of buildings with similar heat load profiles automatically and extract representative patterns showing the characteristics of each group.
2. Customers of interest: we identify customers whose heat load profiles indicate potential problems and require further investigation. In particular, we detect two types of customers, i.e., those deviating significantly from their expected heat load patterns, and those with control strategies determined to be unsuitable for their category.
3. Large-scale evaluation: we present a large-scale analysis of all the buildings in six different customer categories, connected to two DH networks in the south of Sweden. This is the first study analyzing both individual and group behaviors of all the DH customers in an entire system.

The rest of this paper is organized as follows. Section 2 surveys briefly related work on heat load patterns, clustering, and clustering of heat loads. Section 3 first introduces some important concepts and an overview of the data-driven approach, then presents all the steps of the proposed method in detail. Next, Section 4 describes the dataset used, and Section 5 provides the comprehensive results from the analysis of the real-life case study in Sweden. This is followed by a discussion of the results in Section 6. Finally, we conclude in Section 7.

## 2. Related work

In this section, we present related work. First, we give an overview of the related works in the domain of heat-load pattern analysis. Second, we review the state-of-the art concepts related to cluster analysis, which we use in our data-driven approach. Third, a short overview is presented for five previous analyses of the clustering of heat loads in DH systems.

### 2.1. Heat load patterns

Due to the unavailability of high-resolution, hourly, or sub-hourly meter data before the installation of smart meters, the literature on energy analytics in DH is still in its infancy. Therefore, there are not many studies focusing on the analysis of heat load patterns in DH systems.

In [10–12], the heat load patterns were analysed in order to estimate heat load capacities for billing purposes. An approach to separate domestic hot water from space heating using existing heat meters is proposed in [13]. Heat loads were monitored and evaluated in [14] to increase energy efficiency in multi-dwelling buildings.

Energy signature (ES) methods have been used for characterizing heat load behaviors of buildings in multiple studies for purposes ranging from weather corrections [8], to the estimation of heat loss [15–17], to identifying abnormal demands [18]. However, they only reflect individual heat demand as a function of outside temperature over the course of a year. ES methods do not allow profiling of the buildings based on other aspects such as daily behavior, weekend routines, and so forth.

More recent works have targeted applications in peak forecasting [19] and peak shaving [20]. They concern mainly energy conservation via reducing peaks in the daily patterns of load curves, which are 24-h records of the heat loads. However, daily load curves depend heavily on the effects of weather and mostly reflect temporary behaviors rather than regular ones.

Our work is complementary to the studies by Werner and Gadd, which presented a method to analyze heat load patterns manually in [21] and a set of rules for identifying unsuitable behaviors of buildings in [22]. We leverage those works as prior domain knowledge and incorporate some of the concepts that they introduced into the automatic discovery of the heat load patterns. Furthermore, we formalize the individual heat load behavior and group behavior separately in this study, while previous works have not made a clear distinction between these concepts in heat load pattern perspectives.

### 2.2. Clustering

Clustering is the task of organizing data in such a way that similar objects are placed into related or homogeneous groups without prior knowledge of the groups' definitions [23]. It is one of the most popular methods in exploratory data analysis, as it identifies structures in an unlabeled dataset by organizing data into groups that are objectively similar [24]. Numerous techniques have been proposed in the literature for finding clusters in different types of data. In this study, we are concerned mostly with time-series clustering techniques since individual heat load behaviors are used to represent the DH customers as a function of time.

Time-series clustering is a special case of cluster analysis that has been used in many scientific areas to discover interesting patterns in time-series datasets such as smart meter datasets. Many time-series clustering algorithms have been proposed in the literature. There are generally three different approaches to cluster time-series: the feature-based, model-based, and shape-based approaches [23].

In the feature-based and model-based approaches, the raw time series are either converted into a feature vector of lower dimension or transformed into model parameters so that classical clustering methods can be applied [25]. However, feature-based and model-based techniques can lead to loss of information, and they present drawbacks such as the application-dependence of the feature selection or problems associated with parametric modeling [26].

The shape-based approach takes mostly traditional clustering algorithms and modifies the similarity measure to match the shapes of two time series as well as possible. This approach has also been labeled as a raw-data-based approach because it typically works directly with the raw time series data, in contrast to the feature- and model-based approaches. Shape-based algorithms usually employ conventional clustering methods which are compatible with static data, while their distance/similarity measure has been modified with a measure appropriate for time series.

Shape-based methods are highly dependent on the similarity measure [27]. The most well-known distance measure in data mining literature is the Euclidean distance. While the Euclidean distance works well in general for time series clustering tasks, it does not always produce accurate results if data are even shifted slightly along the timeline, and it is very sensitive to noise.

On the other hand, the Dynamic Time Warping (DTW) [28] distance measure and its variances are more suitable for most time series data mining tasks due to its improved alignment based on shape [29]. However, these approaches are computationally expensive and inefficient for time-series averaging tasks. Further, the shape of the cluster center does not represent the characteristics of sequences accurately within the same cluster [30,31].

K-shape clustering [32] is a centroid-based clustering algorithm that is quite similar to the well-known k-means algorithm. However, the k-shape introduces two novel components: (1) shape-based distance (SBD) for dissimilarity measurements, and (2) a time-series shape extraction method for centroid computation that differs from that of k-means. The first one allows the similarities of time-series sequences to be measured based on their shape characteristics, while the second component helps to extract the representative pattern that summarizes the behavior of the cluster.

Extensive evaluation of 48 time series datasets has shown that the SBD measure outperforms Euclidean distance (ED) [33] as well as other state-of-the-art partitioned, hierarchical, and spectral clustering approaches significantly. It also achieves results similar to those of constrained Dynamic Time Warping (cDTW) [34], which is the best-performing distance measure [35], without requiring any parameter tuning and with a much faster performance.

### 2.3. Clustering of heat loads

Recently, five papers have been published about the clustering of heat load patterns in DH systems. An overview concerning the main features of these works is provided in Table 1. These analyses were performed for buildings located in Trondheim [36], Aarhus [37] and [38], Copenhagen [39], and Tianjin [40].

The cluster analyses consider mainly one specific customer category, such as single-family houses or a group of specific service sector buildings. The analysed time periods vary from just one month to up to almost seven years. The magnitude of the analysed objects is not always reported, but the most extensive study included 1.1 million square meters and an annual heat demand of 360 TJ. The common denominator for all five analyses is that they focused on daily heat load patterns. None of these analyses used weekly heat load patterns.

## 3. Methodology

This study aims to provide a data-driven approach, which enables automatic extraction of novel, useful knowledge to better understand the behaviors of complex district heating systems. In general, the problem that we are trying to solve can be seen as an instance of knowledge discovery in databases (KDD) [41]. Specifically, our work is inspired by the nine-stages KDD process model described by Fayyad et al. in [42].

The first stage of this model is developing an understanding of the application domain and the relevant prior knowledge. We use previous works presented in [21] and in [22] to define the key concepts of the problem and specify the goals in the district-heating domain. Following the rest of stages in the model, we formalize a data-driven method for heat load pattern discovery that involves explicit participation of the domain expert. In this chapter, we present the overall knowledge discovery process by first introducing the key concepts that are identified using prior knowledge and then describing the details of the data-driven approach.

Table 1
Overview providing information about five earlier analyses concerning clustering of heat load patterns in district heating systems.

| Location | Buildings | Time period | Floor space, million m² | Annual heat demand, TJ | Clustering algorithm | Reference |
|---|---|---|---|---|---|---|
| Trondheim, Norway | 19 educational buildings | 16 months during two heating seasons, 2011–13 | 0.2 | 72 | Partitioning Around Medoids (PAM) | [36] (Ma, Yan, & Nord, 2017) |
| Aarhus A, Denmark | 8293 single-family buildings | Varying, but the longest was 81 months | 1.1 (estimated) | 360 (estimated) | K-Means | [37] (Gianniou, Liu, Heller, Nielsen, & Rode, 2018) |
| Aarhus B, Denmark | 49 group substations for local networks | One month (January 2017) | Not available | Not available | K-Means | [38] (Tureczek, Nielsen, Madsen, & Brun, 2019) |
| Copenhagen, Denmark | 97 buildings | One month (31 days in March-April) | Not available | Not available | K-Means based online clustering | [39] (Le Ray & Pinson, 2019) |
| Tianjin, China | One group substation for 6 office buildings | Four months (one heating season) | 0.24 | Not available | Gaussian Mixture Model (GMM) | [40] (Lu et al., 2019) |

### 3.1. Concepts

We introduce some important definitions for the problem of heat load pattern discovery.

**Definition 1** (*Heat Load*). Heat load is the quantity of heat per unit time that must be supplied in order to meet the demand in a building.

**Definition 2** (*Heat Load Profile*). A heat load profile is the average hourly heat load of a single building as a function of time.

Given a building $b$ and seasons $S = s_1, s_2, s_3, s_4$, let $M^s \in \mathbb{R}^{h*w}$ be a matrix of hourly heat measurements of $b$ recorded by a single meter, where $h = 24 * 7 = 168$ is the number of hours in a week and $w$ is the number of weeks in season $s$. Heat load profile $\hat{P} = \{A_1, A_2, A_3, A_4\}$ is the set of vectors derived from the four seasons $S$, where $A_s = [a_1^s, a_2^s, \cdots, a_{168}^s]$ is a vector of averages of columns such that

$$a_i^s = \frac{1}{w} \sum_{j=0}^{j=w} M_{ij}^s$$

We define the four seasons in a calendar year as winter (12 weeks of December, January and February), early spring and late autumn (18 weeks of March, April, October and November), late spring and early autumn (9 weeks of May and September) and summer (13 weeks of June, July and August).

Intuitively, heat load profiles capture the recurrent behavior of a building over the whole year with the hourly variations during the day, the changes across weekdays and seasonal differences (Fig. 1).

In this work, our goal is to extract the most typical behaviors in a DH network and represent them as a set of patterns. Clustering aims to group similar objects, while cluster centroids correspond to the average behavior in each group.

Let $N_{\hat{P}} = \{\hat{P}_1, \hat{P}_2, \cdots, \hat{P}_n\}$ be a set of heat load profiles in a DH network. We divide $N_{\hat{P}}$ into $k$ different clusters such that $C = (C_1, C_2, ..., C_k)$, where $C_k \subset N_{\hat{P}}$ and $C_i \cap C_j = \emptyset$. We define $p_i$, a *heat load pattern*, as the centroid of a cluster $C_i$.

**Definition 3** (*Heat Load Pattern*). A heat load pattern is the representation of the central behavior in a group of buildings.

Intuitively, clustering heat load profiles and extracting cluster centroids provides a set of heat load patterns that capture the most typical behaviors in a DH network.

### 3.2. Method overview

In this section, we describe the details of our data-driven approach. This approach is outlined in Fig. 2, and it involves three major steps: (1) data preprocessing, (2) clustering and pattern discovery, and (3) visual exploration. In the first step, the data is cleaned, transformed, and normalized. In the second step, k-shape clustering is performed to group customers with similar heat load behaviors. Abnormal heat load profiles, which do not conform to the behavior exhibited by any group, are detected and removed. Clusters are re-computed without the presence of those buildings, and heat load patterns are extracted. Finally, in the third step, heat load patterns are inspected visually and evaluated qualitatively by the expert. Control strategies are assigned to clusters according to the characteristics of their heat load patterns

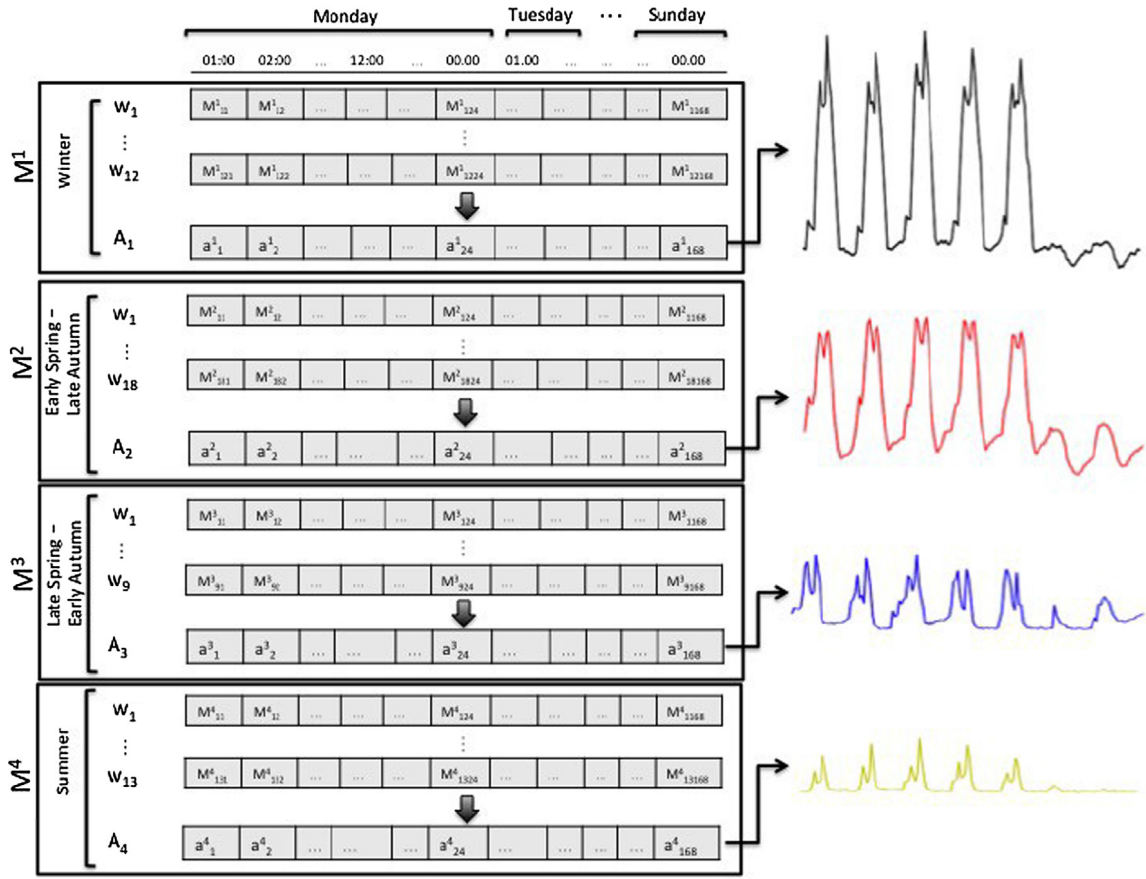

**Fig. 1.** An example showing how to extract a heat load profile. In each week $w$ in matrix $M^s$, there are $24 * 7$ heat load measurements, $\{M_{w1}, M_{w2}, \ldots, M_{w168}\}$. The average weekly heat loads of four seasons form the heat load profile. Then, the heat load profiles are concatenated to single sequence and z-normalized for clustering.

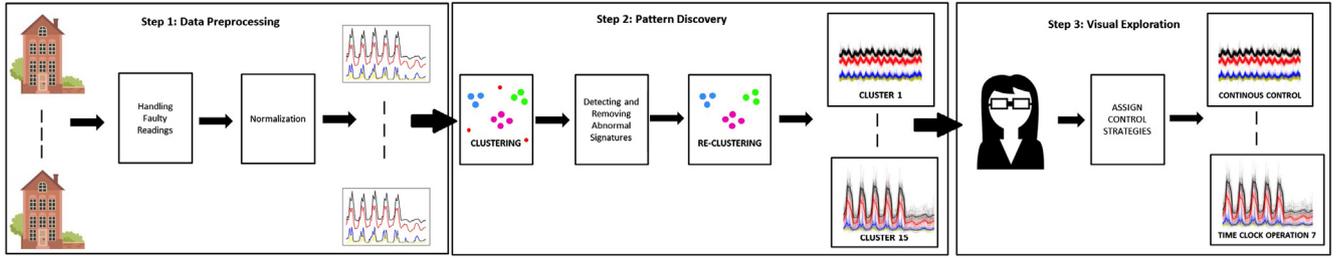

**Fig. 2.** Overview of the three steps in the applied method: data preprocessing (Step 1), pattern discovery (Step 2), and visual exploration (Step 3).

### 3.3. Data preprocessing (step 1)

Starting on 1 January 2015, the Swedish government began requiring DH companies to measure customer heat consumption and charge them accordingly. Therefore, all substations in the Swedish DH systems today are equipped with smart meter devices, which measure the heat used by customers. In Sweden, the data from the meter devices is collected automatically, not manually. The data delivered to a DH company may deviate from the real values due to connection problems. These errors in measurement appear quite frequently in data generated for analysis and visualization purposes.

One of the typical measurement issues is a sudden jump in the heat load. An error can occur if the meter device is not working accurately or if it is replaced by a new device on which the consumption has not been reset correctly. We use median absolute deviation (MAD)-based estimation to detect such extreme values. Heat loads more than five MAD away from the local median are identified as jumps. These values are then corrected by linear interpolation.

Connection problems in the meter device often result in missing values. Buildings with missing sensor readings for more than two consecutive days or more than thirty days in total were excluded from the analysis. Missing values in other cases are filled in the by linear interpolation of surrounding values. Poorly functioning meter devices can also cause repeated measurements. In those cases, meter readings do not change over a certain period. We identified buildings whose values were identical for two consecutive days and excluded them from the study.

After data cleaning, we extract heat load profiles of the buildings to model individual heat load behaviors in the network. Fig. 1 illustrates the computation of each heat load profile, where $\{M_{w1}, M_{w2},…, M_{w168}\}$ corresponds to one week of hourly (24 * 7) heat load measurements in week $w$. Every heat load profile is first merged into a single unified sequence to be used in the clustering process. After clustering, they are converted back to their original format for better visualization.

The last step of the data prepossessing is normalization of the heat load profiles, since the variation in the annual heat demands is high among the customers. This process is essential for the clustering algorithm. As a distance measure, the clustering algorithm uses a normalized version of the cross-correlation measure to consider the shapes of time series. This method is sensitive to scale and requires appropriate normalization to achieve scale invariance. Therefore, all heat load profiles are normalized by z-normalization $\left(z = \frac{x-\mu}{\sigma}\right)$ before the clustering step.

### 3.4. Heat load pattern discovery (step 2)

After data preprocessing, we apply clustering analysis to group similar heat load profiles and extract representative heat load patterns. Heat load profiles reflect how heat is used in an individual building over a year by containing information on changes during the day, differences among weekdays, and seasonal variations. Therefore, it is essential to consider the shape characteristics of these profiles, i.e., the timing and magnitude of its peaks in the clustering process. For this purpose, we apply the k-shape algorithm [32], which is a centroid-based clustering algorithm that can capture the similarities in the shapes of time-series sequences. The k-shape clustering algorithm consists of two main components: (1) a shape-based distance measure, and (2) a time series shape extraction method.

To capture the similarity in the shapes of time-series sequences, k-shape uses SBD, which is based on a normalized cross correlation (NCC). Considering two time series sequences $\vec{x} = (x_1,\cdots,x_m)$ and $\vec{y} = (y_1,\cdots,y_m)$, SBD can be calculated by finding the position $w$ which maximizes $NCC_w(\vec{x}, \vec{y})$ when sliding $\vec{x}$ over $\vec{y}$:

$$SBD(\vec{x}, \vec{y}) = 1 - max_w(NCC_w(\vec{x}, \vec{y})) \quad (1)$$

Cross correlation is computed using Fast Fourier Transformation to reduce the computational complexity of (1) and normalized using the 5th geometric mean of the autocorrelation of each individual heat load profile.

As defined in Section 2.2, a heat load pattern represents a group of heat load profiles and is defined as the cluster centroid. The second component of k-shape clustering deals with computing cluster centroids, where a centroid summarizes the average shape of its cluster. The k-shape algorithm looks at the centroid computation task as a Steiner sequence problem [43] where the objective is to find the minimizer $\vec{\mu_k}^*$ of the sum of squared distances to all other data points.

However, cross-correlation (NCC) which k-shape employs for SBD in (1) measures similarities rather than dissimilarities (distances) of time-series. Therefore, the optimization used to compute centroid $\vec{\mu_k}^*$ is formulated by finding the maximizer of the sum of squared similarities to all other heat load profiles:

$$\vec{\mu_k}^* = \underset{\vec{\mu_k}}{argmax} \sum_{\vec{x_i} \in p_k} NCC_c(\vec{x_i}, \vec{\mu_k})^2 \quad (2)$$

where $p_k$ is the $k$th cluster and $\vec{\mu_k}$ is the initial centroid for the $k$th cluster.

K-shape is a centroid-based algorithm similar to k-means, which means it is similarly sensitive to outliers. Heat load patterns and cluster qualities can be affected by the presence of outliers, i.e., abnormal heat load profiles. Therefore, after removing profiles that are detected as abnormal, we apply the clustering process again to obtain final heat load patterns.

### 3.5. Detecting abnormal heat load profiles

The classical assumption in unsupervised anomaly detection is that anomalies are the sample points which deviate so much from the other sample points as to arouse suspicions that a different mechanism generated them [44]. Based on this assumption, our method quantifies the abnormality of a heat load profile based on its similarity to the cluster to which it belongs. Therefore, we treat distances to the cluster centroids as a measure of abnormality, which means that the more a profile is dissimilar to its heat load pattern (centroid), the more likely it is to be abnormal. In each cluster, we need to estimate a threshold to separate abnormal heat load profiles from the regular ones. An abnormality threshold for each cluster is set to be three standard deviations ($3\sigma$) greater than the mean distance between the cluster members and the centroid. This threshold is derived from Cantelli's inequality [45] with an upper bound of 10% on the false positive rate.

For each $i = 1,2,3,...,k$, let $D_i$ be the vector of distances to centroids in cluster $C_i$ and $\mu_i$ and $\sigma_i$ be mean and standard deviation of $D_i$, then the abnormal profiles are determined as follows:

$$f(x_i) = \begin{cases} abnormal, & \text{if } \eta(d_i, a) \geq 0 \\ nominal, & \text{otherwise} \end{cases} \forall x_i \in C_i, d_i \in D_i \quad (3)$$

where $\eta(d_i, a) = d_i - \mu - a\sigma$ and $a = 3$.

It is important to note that our general hypothesis is based on regular (normal) heat load profiles exhibiting clustering behavior, while abnormal profiles do not conform well to those clusters. This method is not suitable if heat load behaviors in a DH network are sparse and cannot be grouped properly.

### 3.6. Visual inspection of identified clusters (step 3)

In this step, all extracted patterns and their profiles are visualized in a fashion that can help the domain expert to examine them. The visualization should help the expert grasp the typical behavior of each group quickly. Moreover, it should reflect the diversity of individual behaviors within the group so that the cluster quality can be validated quickly.

To this end, we will visualize clusters by plotting heat load patterns with opaque colors and heat load profiles of the buildings with transparency. With this type of visualization, it is also possible to observe the variation among all cluster members and how densely they are populated.

At the final step of the visual inspection, the expert assigns control strategies to the clusters. If the heat load pattern of a cluster reflects the characteristics of one control strategy, the cluster and its members are assigned to that strategy.

## 4. Dataset

The dataset used in this study is derived from smart meter readings from buildings connected to the two DH systems in the Helsingborg and Ängelholm municipalities in the south-west of Sweden. The two cities are not separated in the analysis since they are located close to each other and we did not expect them to generate different cluster groups.

In the beginning of 2016, heat was delivered to 13,766 delivery points that decreases to 2804 if all single-family houses are excluded. During the same year, the local DH provider (Öresundskraft) sold 3780 TJ heat in these two urban areas. This corresponds to a heated floor space of 8.3 million square meters at a specific heat demand of 450 MJ/m².

The data set included originally one year of hourly measurements of heat, flow, supply, and return temperatures on the primary side of all substations during 2016. In this study, we only use the heat measurements of the buildings in six defined customer categories: multi-dwelling buildings, commercial buildings, public administration buildings, health and social service buildings, school buildings, and industrial buildings. Hence, the large set of single-family houses has been excluded from the analysis. The total number of buildings for these categories was 2239. Hence, the dataset consisted initially of 19.6 million potential hourly measurements.

## 5. Results

### 5.1. Data preprocessing (step 1)

In the first data preprocessing stage, 854 buildings were excluded since complete 12-month time series could not be obtained for these buildings for 2016. The main reasons for these losses were explained earlier in Section 3.3. The high rejection rate of 38 percent reveals that the supply chain from heat meters to computers has not yet the reliability to automatically supply high-resolution measurements from all customers for data mining purposes. Hereby, 2239–854 = 1385 buildings remained as input for the clustering process.

### 5.2. Clustering of heat loads (step 2)

In this section, we present the results of the clustering process. To determine the optimal number of clusters (k), we used silhouette analysis [46], which is an internal cluster validation technique that quantifies the clustering quality. Silhouette coefficients range between −1 and 1, where a higher value indicates a better clustering quality. As found by [47], an average silhouette between 0.5 and 0.7 suggests reasonable partitioning of the data, while values higher than 0.7 shows excellent separation between clusters.

In our analysis, we have computed average silhouette coefficients with the number of clusters, $k$, varying between 2 and 30. The results suggest that the maximum average silhouette score (0.61) is found when $k = 15$. After removing abnormal profiles, this average silhouette coefficient increases from 0.61 to 0.69, which indicates high cluster quality. In order to further validate the choice of $k = 15$ for our data set, we also employed visual model selection in which the clustering qualities with different values of $k$ are examined by the domain expert. We have observed that choosing $k$ lower than 15 leads to disappearance of some of the novel patterns that we discovered while choosing higher $k$ leads to redundant patterns that look quite similar to each other.

The fifteen identified clusters contained together 1222 buildings with 10.7 million hourly measurements of heat deliveries. The heat load patterns for these clusters are presented in detail in Section 5.4. The clustered buildings had a total annual heat delivery of 1540 TJ, corresponding to 41 percent of all heat deliveries during 2016. This total delivery is equivalent to a total floor space of 3.4 million square meters.

Hence, 1385–1222 = 163 buildings had heat load profiles that did not fit into the fifteen identified clusters. These outliers with unusual heat load patterns are labeled as abnormal heat load profiles, and some examples of these profiles are presented in the next section.

### 5.3. Abnormal heat load profiles (step 2)

Heat can be used in many different ways inside a building depending on customer behaviors and applied control strategies. The profiles that are identified as abnormal by our method can be considered novel and rare individuals. The domain expert investigated all these profiles and determined that they all have significantly different characteristics from the 15 clusters presented later in Section 5.4 and therefore arouse suspicions. The underlying reasons for showing a "suspicious" profile can vary by building. In some cases, a strange profile can be the symptom of an actual fault in the customer substation that requires further root cause analysis. Fig. 3 presents four examples of buildings with abnormal heat load profiles.

The first building (Fig. 3a) shows a strange trend in which demand increases from Monday to Saturday. It also has inconsistent daily variations in which some days have higher night loads while other days do not. Further analysis revealed that this is a building containing a restaurant and nightclub. The nightclub is open on Fridays and Saturdays, which explains the high heat demand on those days. The low heat loads on Sundays also indicate that there are not many customers on that day.

The second building (Fig. 3b), on the other hand, has atypical behavior in terms of increased weekend loads in colder seasons. This building belongs to a graphical design company. The reason behind the high heat load during weekends and nights could be that the building is heated partially by excess heat from machines during the daytime on weekdays. Moreover, return temperature measurements of this building are consistently high over the course of the year [22].

The third building (Fig. 3c) exhibits extremely sharp, irregular afternoon peaks five days of the week. This building contains some locker rooms, as it is next to a football ground. The use of domestic hot water for showers could explain the peaks. However, it is still difficult to interpret why almost nobody takes a shower on Wednesdays and Sundays.

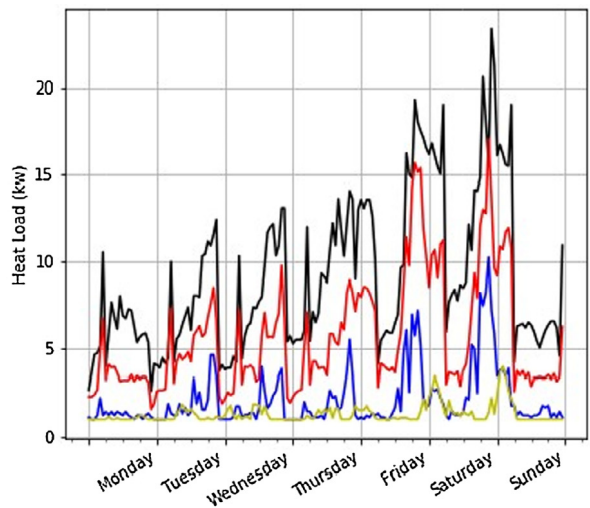

(a) Commercial building

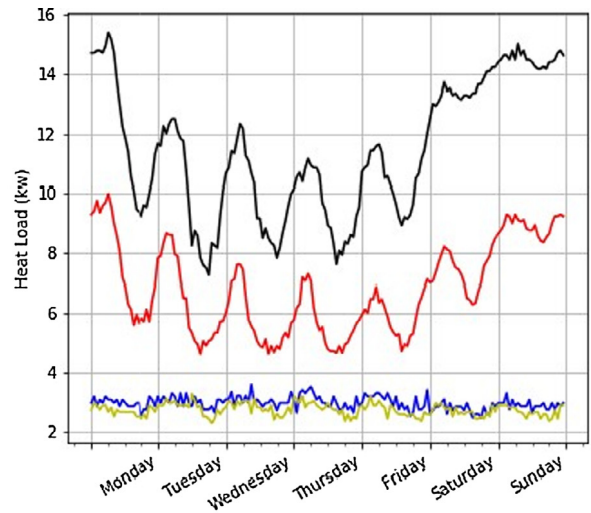

(b) Industrial building

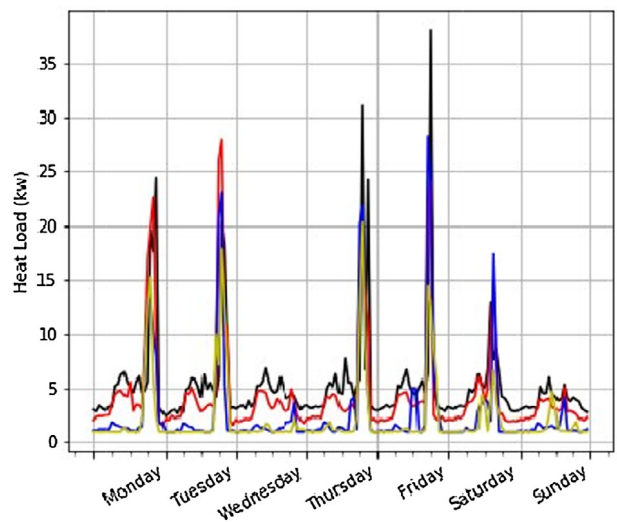

(c) Public administration building

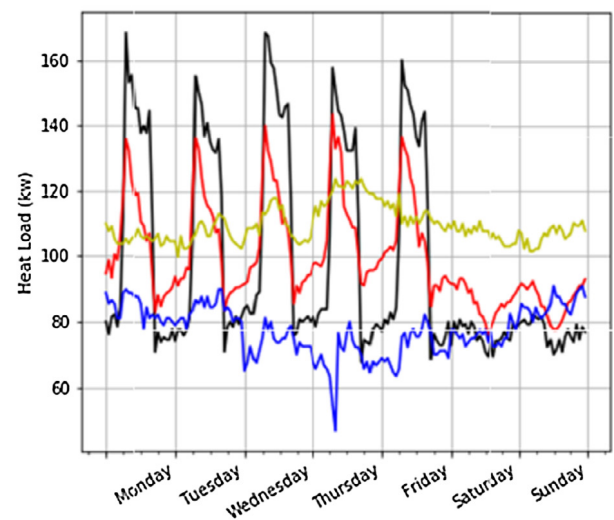

(d) Public admministration building

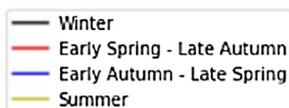

**Fig. 3.** Example abnormal profiles that are selected among 163 profiles identified by our method.

The fourth building (Fig. 3d) is an example of seasonal abnormality where the summer profile is exceptionally high. This building is recorded with unusual heat loads during June and July in 2016 because of a fault occurred in the customer's substation. However, it shows regular behavior of TCO5 control in other seasons. It is important to note that this faulty building would not be detected if we didn't consider summer measurements while profiling individual behaviors.

Those examples show that not all "abnormal" profiles indicate an actual "anomaly" or "fault" in the system. In many other cases, these profiles look much different just because activities and operations in those buildings are rare or unique. Yet, all the 163 profiles with "abnormal" shapes are of special interest in terms of developing more in-depth knowledge about the customers and a better understanding of their heat use.

### 5.4. Identified control strategies (step 3)

Heat load patterns are affected primarily by the control strategies that are applied in the substations for the buildings. Based on earlier domain knowledge [21], we apply four different control strategies in the identification of the fifteen clusters. These control strategies are continuous operation control (COC), night setback control (NSB), time clock operation during the five workdays (TCO5), and time clock operation during all seven days in a week (TCO7).

In COC, the ventilation and radiator systems are running 24 h a day.

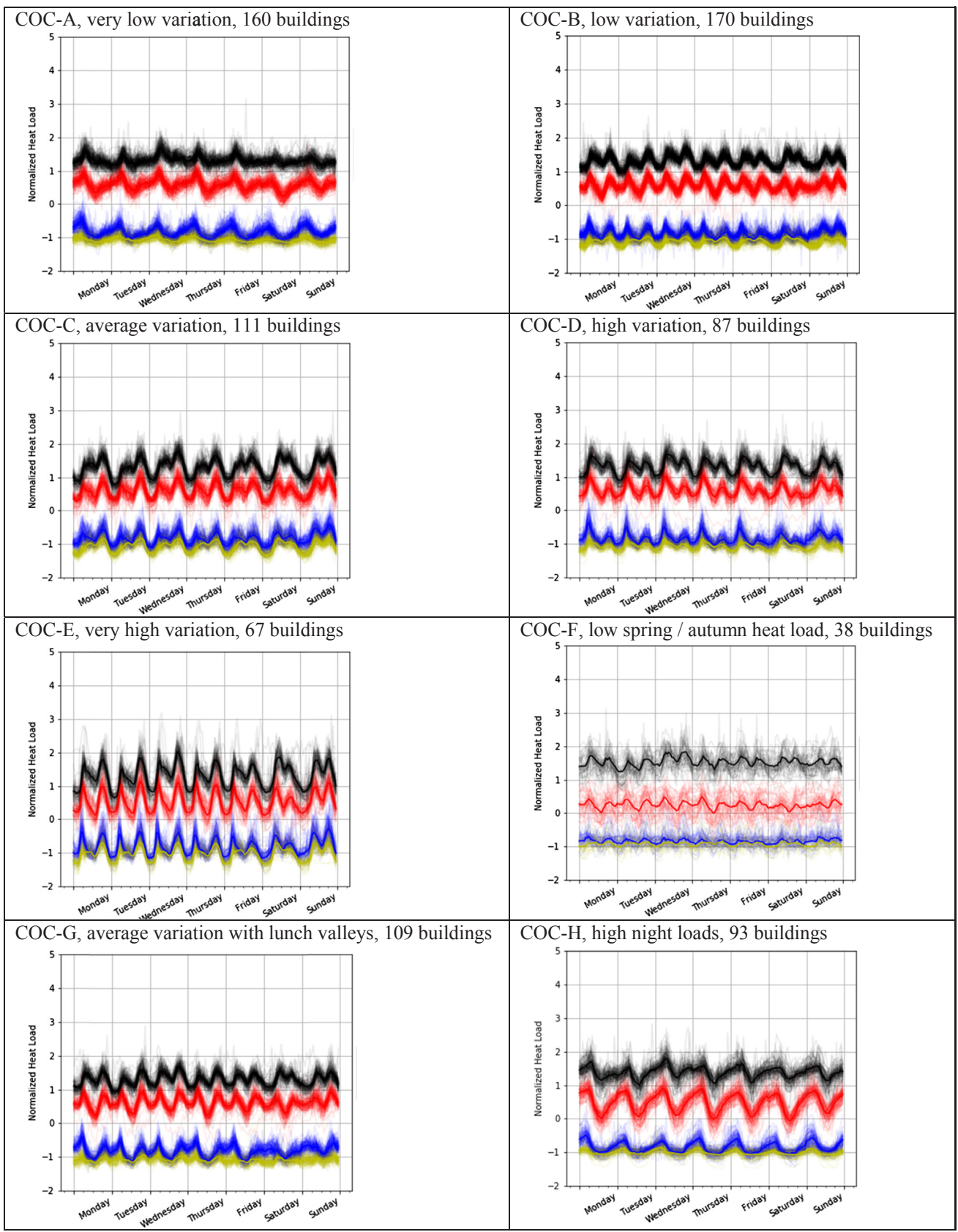

**Fig. 4.** Heat load patterns for the eight cluster groups associated with continuous operation control.

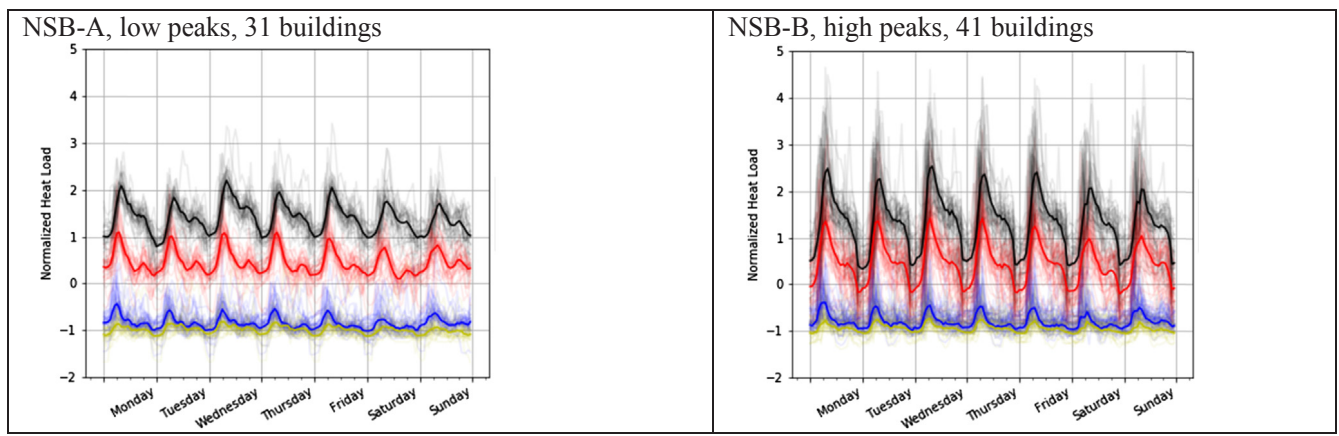

**Fig. 5.** Heat load patterns for the two cluster groups associated with night setback control.

Therefore, it is expected that hot water preparation will cause small variations in the heat load during the day. The NSB lowers the set point for the indoor temperature during the night, leading to lower heat loads during nights, which are followed in the mornings by high peak loads that vanish quite fast. In TCO7, the ventilation system is shut off during the night, resulting in large differences between day and night. The TCO5 operates in a similar fashion, but the ventilation is also completely shut off during weekends.

According to the expert validation of all clusters, eight of them show the main characteristics of COC. They are presented in Fig. 4. The differences in their heat load patterns points out interesting heterogeneity in the behaviors of the buildings controlled with the same strategy.

Furthermore, the domain expert identified two clusters with NSB, which are presented in Fig. 5. Both of them have reduced night loads, which is a typical characteristic of NSB. However, the NSB-B pattern clearly has larger morning peaks in comparison to NSB-A, which could be attributed to a difference in how much the set points for the night setback are reduced.

The remaining five clusters are identified as time clock operation control systems with ventilation that turns on during the daytime and off at night. Two TCO7 cluster groups are reported in Fig. 6, while three TCO5 cluster groups are presented in Fig. 7. The clearest distinction between the two TCO7 clusters is weekend behavior. TCO7-A shows the typical features of TCO7, with similar peaks every day, while TCO7-B has reduced peaks during weekends. This second pattern is probably a mixture of TCO7 and TCO5 systems inside the buildings. Also, according to a previous study [21], it is typical to observe that either there is no significant activity during the weekend in the building or the weekend behavior is similar to the rest of the week.

The three TCO5 variants show various magnitudes of weekday peaks. These magnitudes should differ according the proportion of heat distributed by the ventilation systems. Higher proportions will generate higher peaks.

Concerning seasonal variations in the fifteen cluster groups, many clusters have small differences between the summer and late spring/early autumn seasons. This finding reveals that a higher proportion of the internal heat gains cover a considerable part of the heat demands in the late spring/early autumn season. Hence, more energy-efficient buildings are reducing the length of the heating season. At least in the Helsingborg and Ängelholm locations, May and September should be part of the summer season, while April and October can be merged into the spring/autumn season. Hence, only three different seasons should be used in future cluster analyses.

One important conclusion from these four diagrams is that the cluster analysis did not provide any new heat load patterns beyond the four major control strategies that were defined previously in [21].

Control strategies also play an important role in effective demand-side management and improving building energy performance. In order to improve network governance, it is important to understand how the existing customers operate and control their substations across the entire network. DH companies mostly do not have information about the control strategy that is employed in each building. Our method allows domain experts to determine control strategies for the buildings by exploring the heat load patterns visually on a screen instead of examining thousands of customer installations one-by-one.

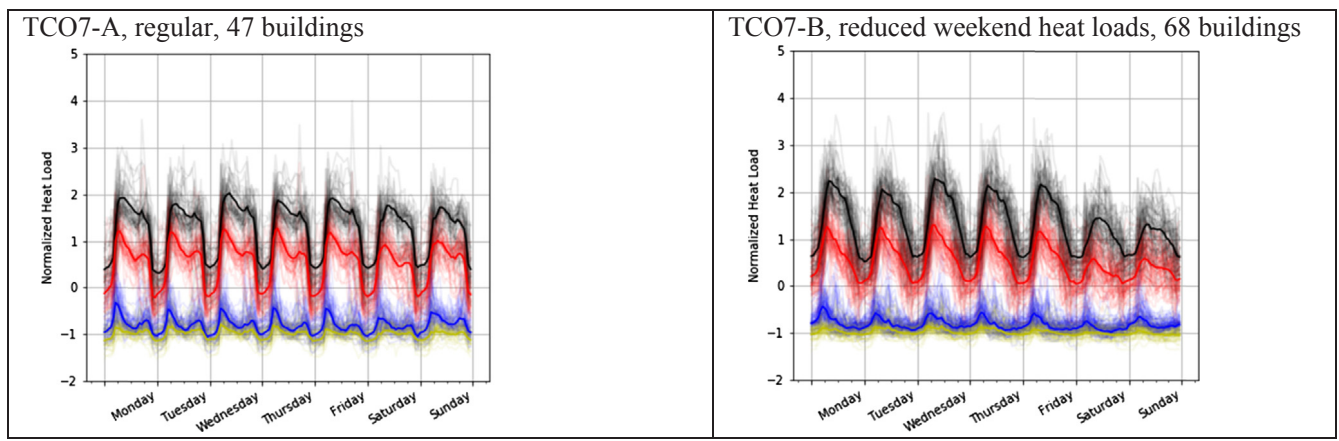

**Fig. 6.** Heat load patterns for the two cluster groups associated with time clock operation during seven days.

E. Calikus, et al.                                                                                                                                              Applied Energy 252 (2019) 113409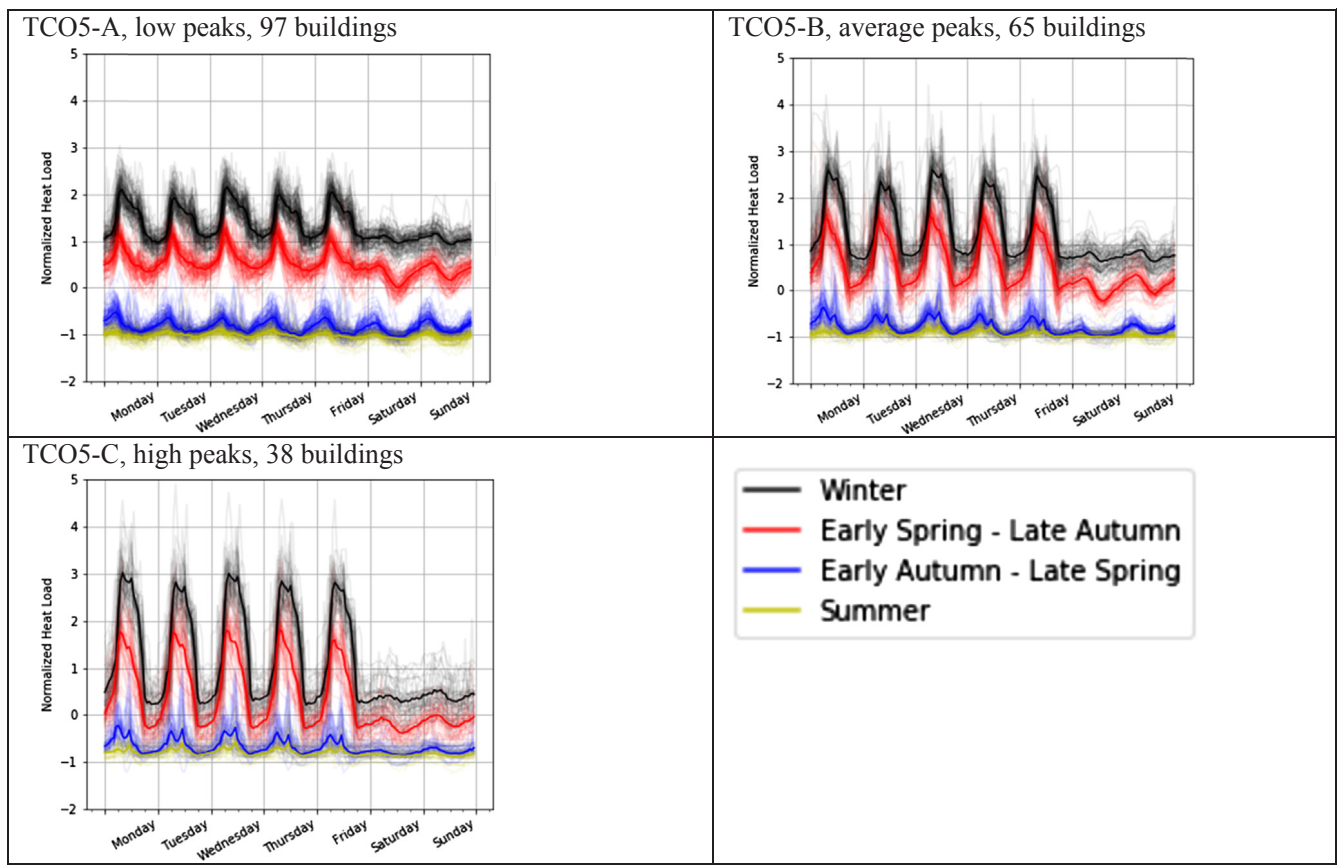

**Fig. 7.** Heat load patterns for the three cluster groups associated with time clock operation during five days.

*5.5. Control strategies versus customer categories*

In this section, the focus is on linking the content in the fifteen cluster groups to the six different customer categories. These links are presented in Fig. 8, while the final overview of the number buildings in various steps of the cluster analysis is presented in Table 2.

COC is the most common control strategy, used on 68% of buildings with 73% of the heat demand. Multi-family buildings dominate the COC group since the final users have a continuous demand for ventilation and heat in order to maintain a constant indoor temperature. Many commercial buildings also use COC. It is also astonishing that 59% of all industrial buildings use COC without any time clock operation at all. Both these customer categories use either the COC-A variant with very low variations or the COC-H variant with high night heat demands. The latter strategy reveals that high daytime internal heat gains probably come from machinery, lighting, or humans.

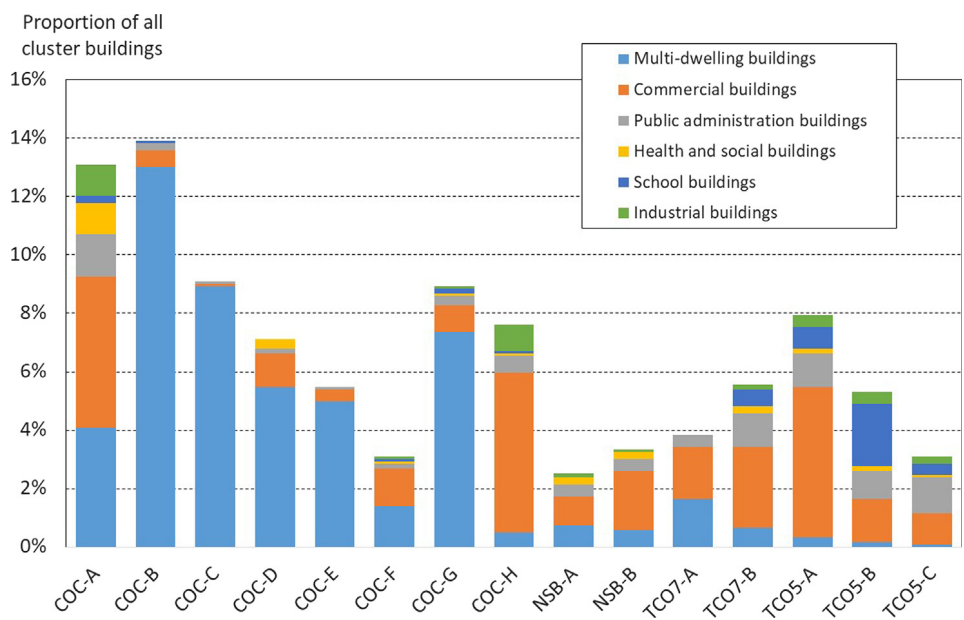

**Fig. 8.** Division of the 1222 clustered buildings with respect to the fifteen cluster groups and the six customer categories.

E. Calikus, et al.                                                                                                                                                      Applied Energy 252 (2019) 113409E. Calikus, et al.                                                                                                                                                      Applied Energy 252 (2019) 113409

**Table 2**
Total overview of the fifteen clusters, plus aggregated groups, buildings with unsuitable control, abnormal heat load profiles, and buildings rejected in the pre-processing stage, with respect to the six customer categories. Information about the annual heat delivery is also provided for the fifteen cluster groups and some aggregated groups.

| Control strategy | Multi-dwelling buildings | Commercial buildings | Public administration buildings | Health and social buildings | School buildings | Industrial buildings | Total buildings | Annual heat delivery, TJ | Time for highest daily peak |
|---|---|---|---|---|---|---|---|---|---|
| COC-A | 50 | 63 | 18 | 13 | 3 | 13 | 160 | 253 | 8:00 AM |
| COC-B | 159 | 7 | 3 | 0 | 1 | 0 | 170 | 234 | 7:00 PM |
| COC-C | 109 | 1 | 1 | 0 | 0 | 0 | 111 | 181 | 7:00 PM |
| COC-D | 67 | 14 | 2 | 4 | 0 | 0 | 87 | 96 | 8:00 AM |
| COC-E | 61 | 5 | 1 | 0 | 0 | 0 | 67 | 70 | 7:00 PM |
| COC-F | 17 | 16 | 2 | 1 | 1 | 1 | 38 | 25 | 8:00 AM |
| COC-G | 90 | 11 | 4 | 1 | 2 | 1 | 109 | 188 | 6:00 PM |
| COC-H | 6 | 67 | 7 | 1 | 1 | 11 | 93 | 73 | 7:00 AM |
| NSB-A | 9 | 12 | 5 | 3 | 0 | 2 | 31 | 45 | 8:00 AM |
| NSB-B | 7 | 25 | 5 | 3 | 0 | 1 | 41 | 27 | 8:00 AM |
| TCO7-A | 20 | 22 | 5 | 0 | 0 | 0 | 47 | 35 | 9:00 AM |
| TCO7-B | 8 | 34 | 14 | 3 | 7 | 2 | 68 | 96 | 8:00 AM |
| TCO5-A | 4 | 63 | 14 | 2 | 9 | 5 | 97 | 125 | 8:00 AM |
| TCO5-B | 2 | 18 | 12 | 2 | 26 | 5 | 65 | 74 | 8:00 AM |
| TCO5-C | 1 | 13 | 15 | 1 | 5 | 3 | 38 | 19 | 8:00 AM |
| Total | 610 | 371 | 108 | 34 | 55 | 44 | 1222 | 1540 | |
| *Aggregated groups:* | | | | | | | | | |
| COC | 559 | 184 | 38 | 20 | 8 | 26 | 835 | 1120 | 73% |
| NSB | 16 | 37 | 10 | 6 | 0 | 3 | 72 | 71 | 5% |
| TCO7 | 28 | 56 | 19 | 3 | 7 | 2 | 115 | 131 | 9% |
| TCO5 | 7 | 94 | 41 | 5 | 40 | 13 | 200 | 218 | 14% |
| Total | 610 | 371 | 108 | 34 | 55 | 44 | 1222 | 1540 | |
| *Thereof unsuitable control:* | | | | | | | | | |
| COC | | 184 | | | | 26 | 210 | | |
| NSB | 16 | 37 | 10 | 6 | 0 | 3 | 72 | | |
| TCO7 | 28 | | | | | | 28 | | |
| TCO5 | 7 | | | | | | 7 | | |
| Total | 51 | 221 | 10 | 6 | 0 | 29 | 317 | 304 | |
| *Abnormal heat load profiles:* | | | | | | | | | |
| Total | 51 | 67 | 24 | 3 | 2 | 16 | 163 | 179 | |
| *Number of buildings for input to the clustering process:* | | | | | | | | | |
| Total | 661 | 438 | 132 | 37 | 57 | 60 | 1385 | 1720 | |
| *Number of buildings rejected in the data pre-processing stage* | | | | | | | | | |
| Total | 411 | 253 | 112 | 17 | 9 | 52 | 854 | | |
| *Number of buildings in all six customer categories* | | | | | | | | | |
| Total | 1072 | 691 | 244 | 54 | 66 | 112 | 2239 | | |

NSB is the least common control strategy, with only 6% of the buildings and 5% of the heat demand. Commercial buildings dominate the NSB group, but it also includes some multi-family buildings that are still using this old and outdated control strategy, originally introduced for inefficient buildings with high heat demands. It was then easy to reduce the indoor temperature quickly in order to reduce the heat demand. This control method is unsuitable for modern and efficient buildings with low heat demands since the indoor temperature cannot be reduced quickly enough during the setback interval.

The TCO7 control strategy is also used primarily by commercial buildings since the ventilation rates can be reduced during nights when no activity occurs. Some multi-family buildings use also this control strategy, especially the TCO7-A variant with its unreduced peaks during weekends. It can be that buildings are labeled as multi-family buildings when they actually are mixed residential and commercial.

The TCO5 control strategy is used for commercial, public administration, and school buildings, especially buildings with no activity during weekends. Concerning the schools, 73% of all school buildings use TCO5, especially the TCO5-B variant. Hence, this customer category has high compliance with a proper control strategy. However, some schools still use COC, leaving some improvement potential in terms of lower heat demands with TCO5.

A concluding remark is that 45% of all service sector and industrial buildings still use COC as a control strategy. This is almost the same proportion as that for the two TCO strategies (46%). An ambition can be that most of these COC buildings can apply TCO7 or TCO5 in the future. Switching control strategies would then provide more space for replacing peak heat loads with daily thermal storage and heat generation during nights. However, one barrier for this improvement could be lack of mechanical ventilation systems in these COC buildings.

*5.6. Unsuitable control*

As mentioned in the previous section, less appropriate control strategies are applied in many buildings that are included in this cluster analysis. According to domain knowledge [22], we consider the following rules to identify unsuitable control strategies:

- Multi-family dwellings that do not have continuous control
- Commercial and industrial buildings that do not have time clock operation
- Any building with night setback control

These three rules are based on three simple conditions for the choice of control strategy for a heated building. First, multi-family buildings have residents that require continuous heat delivery according to the outdoor temperature for maintaining comfortable air rates and indoor temperatures. Second, commercial and industrial buildings should reduce ventilation rates when no activity is occurring. Third, the outdated method of night setback control should not be used in modern buildings with low heat demands since the thermal inertia is high in these buildings. If modern buildings were to still apply night setback control, heat delivery will be directed towards peak times when fossil fuels are often used in peak boiler plants. Hence, elimination of night setback control will reduce carbon dioxide emissions.

According to Table 2, 317 buildings have been identified as buildings with potentially unsuitable control strategies. This corresponds to 20% of the heat demand in all buildings belonging to the fifteen clusters. However, only manual inspections done building by building can reveal the real occurrence of unsuitable control strategies in these 317 buildings.

## 6. Discussions

According to [21], multi-dwelling buildings are assumed to show relatively homogeneous behavior and are expected to exhibit COC. Thus, it is no surprise that the majority of the multi-dwelling buildings in this study are assigned to COC. Yet, a considerable number of buildings had other control strategies. Multi-dwellings with time clock operation were not discovered in previous studies. The reason for such behavior is that some multi-dwellings can also contain restaurants or offices that are limited to daytime activities and have heat load patterns with time clock operation control of ventilation and low domestic hot water use.

Health and social service buildings, as well as commercial buildings, are heterogeneous with regards to heat demand behaviors. In the category of health and social service buildings, there are buildings, such as hospitals, that have 24-h activity patterns similar to those of multi-dwelling buildings. On the other hand, there are also offices within this category that have just time-clock operation controls. Commercial buildings also house different customers, some of which exhibit 24-h activity with domestic water use patterns, such as hotels, and some of which exhibit only daytime activity, such as trading companies, restaurants, and amusement and recreational services. Even though the control strategies used by these two categories are much more diverse in comparison to the strategies used by multi-dwelling buildings, both categories still use predominantly COC control strategies.

Public administration and school buildings are the only customer categories for which COC is not the most frequently used strategy. This outcome is expected, considering that most of the buildings in these categories are municipal buildings with daytime activities only. The school buildings are strongly consistent since very few of them are not controlled with time clock operation controls. The COC rate is much higher for public administration buildings because some of the buildings in this category remain active for 24 h a day, as in, for example, service buildings for seniors.

Future benefits of automatic discovery of heat load patterns include more efficient DH systems with respect to heat demands, temperature levels, and carbon dioxide emissions. By reducing unsuitable heat use, customers can cut their heating bills. Reductions in heat demands also provide the future possibility of operating the DH systems at lower temperature levels, reducing the costs for introducing more renewables and heat recycling [48]. The highest carbon dioxide emissions today are associated with peak heat demands, which will decrease considerably when the customer peak demands are reduced.

As a future work, we plan to extend this study and try and compare different models to represent heat load profiles such as daily representations, incorporating outdoor temperature and weather information, etc. Furthermore, we would like to provide deeper analysis of abnormal profiles that are discovered in Section 5.3 and try to model and learn different abnormal behaviors in the context of rare class detection.

## 7. Conclusions

In this work, we study the problem of discovering heat load patterns in DH networks automatically. We argue that there is a need for a data-driven approach and present three contributions from analyzing the heat load behaviors of DH customers. The first contribution is a method that enables the clustering buildings by preserving the shape similarities in their heat load profiles and extracting patterns summarizing the typical behavior in each group. The second is detecting buildings with abnormal heat load profiles, i.e., those that look significantly different from their expected heat load patterns. The third contribution is the identification of buildings with control strategies that are unsuitable for their customer category, based on visual inspection and the domain experts' validation of discovered heat load patterns.

We conduct a novel case study on two district heating networks in the south-west of Sweden. To the best of our knowledge, this is the first large-scale, comprehensive analysis of heat load patterns that provides insights into the heat load behavior of entire networks. Our method captured fifteen common patterns among the heat load profiles of all the buildings in our dataset and discovered many profiles that are significantly different from those patterns. With the help of this study, we gained an insight into how typical and atypical behaviors can look like in district heating networks and revealed the limitations of the previous knowledge. We have found that heat load behaviors can vary even if the customer substations are controlled in the same manner. Furthermore, we have shown that buildings with different customer categories often behave quite similarly, while the ones within the same category can behave differently. Therefore, we can conclude that neither the current customer categories (in Sweden, at least) nor the existing control strategies are sufficient for the categorization of buildings in DH. We believe that our approach has high potential to serve this purpose in practice since it is automatic, can discover knowledge which was previously not known, and can deal with large-scale data.

## Acknowledgements

This research is supported by the Swedish Knowledge Foundation (KK-stiftelsen) with Grant No. 20160103. We also thank Öresundskraft for sharing the smart-meter dataset that was used in this study. The fourth and the fifth authors have acted as domain experts concerning district heating in this work.

## Appendix. The pseudocode for the method described in Section 3.2

**Algorithm 1:** Heat Load Pattern Discovery

**Input** : Dataset $D$;
     Number of clusters $k$;
**Output:** Heat load patterns $P$

$D \leftarrow DataCleaning(D)$ ;   ▷ Data cleaning (Section 3.3)
$\hat{P} \leftarrow ExtractHeatLoadProfiles(D)$ ;   ▷ Extracting heat load profiles (Section 3.3)
$\hat{P}_{normalized} \leftarrow ZNormalize(\hat{P})$ ;   ▷ Normalization (Section 3.3)
$C \leftarrow KShape(\hat{P}_{normalized}, k)$;   ▷ Clustering heat load profiles (Section 3.4)
$\hat{P}_{abnormals} \leftarrow DetectAbnormalProfiles(C)$;   ▷ Detecting abnormal heat load profiles (Section 3.5)
$\hat{P}_{nominals} \leftarrow RemoveAbnormalProfiles(\hat{P}_{normalized}, \hat{P}_{abnormals})$;   ▷ Removing abnormal heat load profiles
$C_{final} \leftarrow KShape(\hat{P}_{nominals}, k)$ ;   ▷ Re-clustering heat load profiles (Section 3.4)
$P \leftarrow GetCentroids(C_{final})$ ;   ▷ Get final cluster centroids
**return** $P$ ;   ▷ Return cluster centroids as heat load patterns

---

**Procedure** DetectAbnormalProfiles(Clusters C)

**Input:** Clusters $C$;
**Output:** Abnormal profiles $\hat{P}_{abnormals}$

**for** $c_i \in C$ **do**
  $dists \leftarrow GetCentroidDistances(c_i)$ ;
  $\mu \leftarrow mean(dists)$ ;
  $\sigma \leftarrow std(dists)$;
  $th \leftarrow \mu + 3 * \sigma$ ;
  **for** $d \in dists$ **do**
   **if** $th < d$ **then**
    $\hat{P}_{abnormals} \leftarrow \hat{P}_{abnormals}.add(GetHeatLoadProfile(c_i, d))$;
   **end if**
  **end for**
**end for**
**return** $\hat{P}_{abnormals}$ ;